\documentclass[aps,superscriptaddress,showpacs,floatfix]{revtex4}
\usepackage{graphicx}
\usepackage{dcolumn}
\usepackage{bm}
\usepackage{CJK}

\begin{document}

\begin{CJK*}{GBK}{song}

\title{Three-body force effect on nucleon momentum distributions in asymmetric nuclear matter within the framework of the
extended BHF approach}

\author{Peng Yin}
\affiliation{Institute of Modern Physics, Chinese Academy of
Sciences, Lanzhou 730000, China} \affiliation{University
of Chinese Academy of Sciences, Beijing, 100049, China}
\author{Jian-Yang Li}
\affiliation{Institute of Modern Physics, Chinese Academy of
Sciences, Lanzhou 730000, China}
\author{Pei Wang}
\affiliation{Institute of Modern Physics, Chinese Academy of
Sciences, Lanzhou 730000, China}
\affiliation{University
of Chinese Academy of Sciences, Beijing, 100049, China}
\author{Wei Zuo \footnote{Corresponding author: zuowei@impcas.ac.cn}}
\affiliation{Institute of Modern Physics, Chinese Academy of
Sciences, Lanzhou 730000, China}

\begin{abstract}
We have investigated the three-body force (TBF) effect on the neutron and proton momentum distributions
in asymmetric nuclear matter within the framework of the extended Brueckner-Hartree-Fock approach by adopting
the $AV18$ two-body interaction plus a microscopic TBF.
In asymmetric nuclear matter, it is shown that the neutron and proton momentum distributions become
different from their common distribution in symmetric nuclear matter.
The predicted depletion of the proton hole states increases while the neutron one decreases as a function of
isospin-asymmetry. The TBF effect on the neutron and proton momentum distributions turns out to be negligibly weak at low
densities around and below the normal nuclear density.
The TBF effect is found to become sizable only at high densities well above the saturation density,
and inclusion of the TBF leads to an overall enhancement of the depletion
of the neutron and proton Fermi seas.

\end{abstract}

\pacs{21.60.De, 21.65.Cd, 21.30.Fe} \maketitle

\section{Introduction}

To determine reliably the properties of isospin asymmetric nuclear matter is a challenge
in nuclear physics and nuclear astrophysics~\cite{ba1,bombaci:1991,Dieperink:2003,djm}. The nucleon momentum distribution
in nuclear matter is one of the most important properties of nuclear matter.
Many-body correlations induced by nucleon-nucleon ($NN$) interactions among nucleons
play a significant role in a nuclear many-body system, which make the system much
more complicated and have more plentiful properties than a non-interacting Fermi
system. For example, the effect of the short-range correlations may lead to the depletion of the
 nucleon momentum distribution below the Fermi momentum and the population above
 the Fermi momentum in nuclear matter~\cite{jeukenne:1976}.  The nucleon momentum distribution is
 of a great physical interest since it is closely related to the nature of the underlying $NN$ interaction.
 The depletion of the Fermi sea is expected to be closely related to the hard core and the tensor
 component of the $NN$ interaction~\cite{vonderfecht:1991}. It plays an important role in testing
 the validity of the physical picture of independent particle motion in the mean field theory
 or the standard shell model and serves as a measure of the strength of the dynamical $NN$ correlations induced by
the $NN$ interaction in a nuclear many-body system~\cite{cavedon:1982,pandharipande:1997}.
The study of the nucleon momentum distribution in nuclear matter
may provide desirable information on the depletion of the deeply bound states inside finite nuclei and
is expected to be important for understanding the structure of finite nuclei.
Experimentally, the effects of $NN$ correlations and the nuclear depletion of the Fermi sea can be investigated by
the ($e,e'p$), ($e,e'NN$), and proton induced knock-out reactions~\cite{ramos:1989,dickhoff:1992,dickhoff:2004}.
The related measurements have been reported
continually~\cite{mitt:1990,lapikas:1999,starink:2000,batenburg:2001,rohe:2004,niyazov:2004,
benmokhtar:2005,aclander:1999,onderwater:1998,reley:2008,subedi:2008} and definite evidence of the
short-range $NN$ correlations has been observed in these experiments. The
analysis of the ($e,e'p$) reactions on $^{208}$Pb at NIKHEF has indicated that a depletion of $15\% - 20\%$ for the deeply bound proton states is required for describing the measured coincidence cross sections~\cite{batenburg:2001}.
Recent experiments on the two-nucleon knock-out reactions have shown that nucleons can form short-range correlated pairs with large
relative momentum and small center-of-mass momentum. A strong enhancement of the neutron-proton ($np$) short-range correlations over the proton-proton ($pp$) correlations has been observed at JLab~\cite{subedi:2008} due to the dominate role played by the short-range tensor components of the $NN$ interactions in generating the $NN$ correlations~\cite{schiavilla:2007}, and
may have significant implications for neutron star physics~\cite{frankfurt:2008}.

The short-range correlations and the nucleon momentum distribution in
 nuclear matter have been investigated extensively by using various theoretical approaches, such as
 the extended Brueckner-Hartree-Fock (BHF) method \cite{sartor:1980,grange:1987,jaminon,baldo:1990,baldo:1991,mahaux:1993,hassaneen:2004},
 the Green function theory \cite{muther:1995,alm:1996,dewulf:2003,frick:2005,rios:2009},
 the in-medium T-matrix method \cite{bozek:2002,soma:2008}, the variational Monte Carlo approach \cite{schiavilla:2007}, and
 the correlated basis function approach \cite{fantoni:1984,benhar:1990}. The predicted depletion of the Fermi sea
by adopting different theoretical approaches has been shown to be slightly larger than
15\% \cite{grange:1987,baldo:1991,fantoni:1984,benhar:1990}.
In Ref.~\cite{baldo:1991}, the authors have calculated the nucleon
 momentum distribution and quasiparticle strength in symmetric nuclear matter
 in the framework of the Brueckner-Bethe-Goldstone theory by including high-order
 contributions in the hole-line expansion of the mass operator, and a
 good agreement between the calculated quasiparticle strength and the experimental data~\cite{mitt:1990} has been shown.
In Ref. \cite{hassaneen:2004}, the neutron and proton occupation probabilities, averaged below their respective Fermi seas in
neutron-rich matter at the saturation density, have been predicted.
In more recent papers \cite{frick:2005,rios:2009}, the nucleon momentum distributions in neutron matter and asymmetric nuclear matter have been
investigated within the framework of the Green function method,
and the isospin-asymmetry dependence of the depletions of the neutron and proton Fermi seas
has been clarified. It has been shown \cite{hassaneen:2004,frick:2005,rios:2009} that increasing the isospin asymmetry leads to
an increasingly larger depletion of the proton hole-states than that of the
neutron hole states in asymmetric nuclear matter.
 In the present paper, we shall
extend the previous investigation of Ref.~\cite{baldo:1991} to asymmetric nuclear matter and
investigate the isospin-asymmetry dependence of the neutron and proton momentum distributions within the
extended Brueckner-Hartree-Fock (EBHF) method. Particulary, we will concentrate on the TBF effect on the
 momentum distributions and their isospin dependence in asymmetric nuclear matter especially at suprasaturation
 densities.

The present paper is organized as follows. In the next section, we
give a brief review of the adopted theoretical approaches including the EBHF theory and the TBF model.
In Sec. III, the calculated results will be reported and discussed. Finally, a summary is given in Sec. IV.

\section{Theoretical Approaches}

The present calculations are based on the extended BHF approach for asymmetric nuclear
matter~\cite{zuo:1999}. The extension of the BHF scheme to
include microscopic three-body forces can be found in Refs. \cite{grange:1989,zuo:2002a,zuo:2002b}.
Here we simply give a brief review for completeness.
The starting point of the BHF approach is
the reaction $G$-matrix, which satisfies the following isospin
dependent Bethe-Goldstone (BG) equation \cite{day}:
\begin{eqnarray}\label{eq:BG}
G(\rho,\beta;\omega)= V_{NN}+
 V_{NN}\sum\limits_{k_{1}k_{2}}\frac{|k_{1}k_{2}\rangle
Q(k_{1},k_{2})\langle
k_{1}k_{2}|}{\omega-\epsilon(k_{1})-\epsilon(k_{2})}G(\rho,\beta;\omega)
\end{eqnarray}
where $k_i\equiv(\vec k_i,\sigma_i,\tau_i)$ denotes the momentum, and
the $z$-component of spin and isospin of a nucleon, respectively.
$V_{NN}$ is the realistic $NN$ interaction and $\omega$ is the starting energy.
The asymmetry parameter $\beta$ is defined as
$\beta=(\rho_{n}-\rho_{p})/\rho$, where $\rho$, $\rho_n$ and $\rho_p$ denote
the total nucleon, neutron, and proton
number densities, respectively.
The Pauli
operator is defined as $Q(k_{1},k_{2})=[1-n_0(k_{1})][1-n_0(k_{2})]$,
and it prevents two nucleons in intermediate sates from being scattered into their
respective Fermi seas (Pauli blocking effect). Here by $n_0(k)$
we denote the Fermi distribution function which is a step function at zero temperature, i.e.,
$n_0(k)=\theta(k_F-k)$.
The single-particle (s.p.) energy $\epsilon(k)$ is given by:
$ \epsilon(k)=\hbar^{2}k^{2}/(2m)+U(k)$,
where the auxiliary s.p. potential $U_{\rm BHF} (k)$ controls the convergent rate of the
hole-line expansion \cite{day}.
In the present calculation, we adopt the continuous choice for
the auxiliary potential since it provides a much faster convergence of
the hole-line expansion up to high densities than the gap choice \cite{song:1998}.
 Under the continuous choice, the s.p. potential describes physically at the lowest BHF level
the nuclear mean field felt by a nucleon in nuclear medium \cite{lejeune:1978}, and is
calculated as follows:
\begin{eqnarray}\label{eq:UBHF}
U(k)=Re\sum\limits_{k'\leq k_{F}}\langle
kk'|G[\rho,\epsilon(k)+\epsilon(k')]|kk'\rangle_{A} \ ,
\end{eqnarray}
where the subscript $A$ denotes anti-symmetrization of the matrix
elements.

For the realistic $NN$ interaction $V_{NN}$, we adopt the
Argonne $V_{18}$ ($AV18$) two-body interaction \cite{wiringa:1995} plus a
microscopic TBF \cite{zuo:2002a} constructed by using the meson-exchange current
approach \cite{grange:1989}.
In the TBF model adopted here, the most important mesons, i.e., $\pi$, $\rho$, $ \sigma $, and $\omega$ have been considered.
The parameters of the TBF model, i.e., the coupling constants
and the form factors,
have been self-consistently determined to reproduce the $AV18$
two-body force using the one-boson-exchange potential
model and their values can be found in Ref. \cite{zuo:2002a}.
 In our calculation, the TBF
contribution has been included by reducing the TBF to an
equivalently effective two-body interaction according to the
standard scheme as described in Ref. \cite{grange:1989}. In
$r$-space, the equivalent two-body force $V_3^{\rm eff}$ reads
\begin{eqnarray}\label{eq:tbf}
 \langle \vec r_1^{\ \prime} \vec r_2^{\ \prime}| V_3^{\rm eff} |
\vec r_1 \vec r_2 \rangle = \displaystyle
 \frac{1}{4} {\rm Tr} \sum_{n} \int {\rm d}
{\vec r_3} {\rm d} {\vec r_3^{\ \prime}}\phi^*_n(\vec r_3^{\
\prime}) (1-\eta(r_{13}')) (1-\eta(r_{23}')) \nonumber \\
\times W_3(\vec r_1^{\ \prime}\vec r_2^{\ \prime} \vec r_3^{\
\prime}|\vec r_1 \vec r_2 \vec r_3)
 \phi_n(\vec r_3)
(1-\eta(r_{13}))
 (1-\eta(r_{23})).
\end{eqnarray}
In order to calculate the nucleon momentum distribution in nuclear matter with the EBHF approach,
we follow the scheme given in Refs. \cite{jeukenne:1976,baldo:1991} and extend the scheme to asymmetric nuclear matter.
 Within the framework of the Brueckner-Bethe-Goldstone theory, the mass operator can be expanded in a perturbation series
according to the number of hole lines, i.e.,
\begin{equation}
 M^{\tau}(k,\omega) = M^{\tau}_1(k,\omega)+M^{\tau}_2(k,\omega)+M^{\tau}_3(k,\omega)+\cdots
 \end{equation}
where $\tau$ denotes neutron or proton (hereafter we will write out explicitly the isospin index $\tau$).
The mass operator is a complex quantity and the real part of its on-shell value
can be identified with the potential energy felt by a neutron or a proton
in asymmetric nuclear matter. In the expansion of the mass operator, the first-order contribution
$M^{\tau}_1 (k,\omega)$ corresponds to the
standard BHF s.p. potential and the on-shell value of its real part coincides
with the auxiliary potential under the continuous choice given by Eq. (\ref{eq:UBHF}).
The higher-order terms stem from
the density dependence of the effective $G$-matrix.
As shown by Jeukenne {\it et al.} \cite{jeukenne:1976}, in order to predict reliably the s.p. properties within the
Brueckner theory,
one has to go beyond the lowest-order BHF approximation by considering the high-order contributions
in the hole-line expansion of the mass operator.
The second-order
term $M^{\tau}_2$ is called the Pauli rearrangement term and it is induced by the medium dependence of the
$G$-matrix via the Pauli operator in the BG equation \cite{jeukenne:1976,zuo:1999}.
The Pauli rearrangement effect of the
$G$-matrix describes the influence of the ground
state two-hole correlations on the s.p. potential \cite{baldo:1988,baldo:1990}.
The ground state $NN$ correlations have been investigated extensively in
literature \cite{schnell,Czerski:2002,frick}. The Pauli
rearrangement have been shown to play its role mainly in the low
momentum region around and below the Fermi surface where the ground
state two-hole correlations are most effective, and it weakens the
momentum dependence of the s.p. potential especially around the
Fermi surface. The effect of ground
state correlations not only is essential for getting a satisfactory agreement between
the predicted depth of the microscopic BHF s.p. potential and the empirical
value \cite{jeukenne:1976} and for restoring the Hugenholtz-Van Hove
theorem which is destroyed seriously at the lowest BHF approximation \cite{zuo:1999},
but also plays a crucial role in generating a nucleon self-energy to describe
realistically the s.p. strength distribution in nuclear matter and finite nuclei below the
Fermi energy \cite{dickhoff:1992}. According to Refs. \cite{jeukenne:1976,zuo:1999},
the Pauli rearrangement contribution is
calculated as follows from the $G$-matrix:
\begin{eqnarray}\label{eq:M2}
M^{\tau}_2(k,\omega)=\sum_{\tau'}M^{\tau,\tau'}_2=
 \frac{1}{2} \sum_{\tau'}\sum_{\sigma'\vec{k}'>k^{\tau'}_F}\sum_{k_1<k^{\tau}_F}\sum_{k_2<k^{\tau'}_F}
 \frac{|\langle kk'|
 G^{\tau,\tau'}[\epsilon^{\tau}(k_1)+\epsilon^{\tau'}(k_2)]|k_1k_2\rangle|^2}
{\omega+\epsilon^{\tau'}(k')-\epsilon^{\tau}(k_{1})-\epsilon^{\tau'}(k_{2})-i\eta} \ .
\end{eqnarray}
Due to the $NN$ correlations in asymmetric nuclear medium, the neutron and proton Fermi seas are partially depleted,
and consequently the correlated momentum distributions differ from the uncorrelated ones.
The third-order term $M^{\tau}_3$ in the hole-line expansion of mass operator
is called the renormalization contribution and it takes into account the above effect of the depletion of the Fermi seas.
The renormalization term $M^{\tau}_3$ is given by \cite{jeukenne:1976,baldo:1990,zuo:1999}
\begin{equation}
M_3^{\tau}(k,\omega)=
- \sum_{\tau'}\sum_{\vec{h'} \sigma'}\kappa_2^{\tau'}(h')
\langle kh'|G^{\tau\tau'}(\omega + \epsilon^{\tau'}(h'))|kh'\rangle_A \ ,
\label{eq:M3}
\end{equation}
where $h'$ refers to the hole state with momentum smaller
than $k_{\rm F}^{\tau}$, and
\begin{equation}
\kappa_2^{\tau'}(h')=-\left[
\frac{\partial}{\partial\omega}
M_1^{\tau'}(h',\omega)\right]_{\omega=\epsilon^{\tau'}(h')}
\end{equation}
is the depletion of the neutron or proton Fermi
sea at the lowest-order approximation in asymmetric nuclear matter \cite{jeukenne:1976,zuo:1999},
{\it i.e.,}  $\kappa_2^{\tau'}(h')$ is the probability that a neutron or proton hole-state
$(|\vec h'| \le k_F^{\tau'})$ is empty.
Hereafter, we shall use $h$ and $h'$ to denote the s.p. hole states below the Fermi momentum.
As shown in Ref.~\cite{baldo:1991}, it is an satisfactory
approximation to replace in Eq. (\ref{eq:M3}) the depletion coefficient $\kappa_2^{\tau'}(h')$  by its value
at the averaged momentum inside the Fermi sea, i.e., $\kappa^{\tau'}=\kappa_2^{\tau'}(h'=0.75k^{\tau'}_F)$.
We have $M_3^\tau(k,\omega) \approx - \sum_{\tau'} \kappa^{\tau'} M_1^{\tau\tau'}(k,\omega)$. By taking into account
the renormalization term $M^{\tau}_3(k,\omega)$, one may get the {\it renormalized} BHF
approximation for the mass operator~\cite{zuo:1999}, i.e.,
\begin{equation}
 \widetilde{M}_1^\tau(k,\omega) \equiv M^{\tau}_1(k,\omega)+M^{\tau}_3(k,\omega) \approx
\sum_{\tau'} \Big[ 1 - \kappa^{\tau'} \Big] M_1^{\tau\tau'}(k,\omega) \ .
\end{equation}
Similarly, one may consider a renormalization correction
from the four hole-line terms to the
second-order term $M_2^{\tau}$ in order to take
into account the fact that the hole-state $k_2$ in
Eq.~(\ref{eq:M2}) is partially
empty ( see also Ref.~\cite{jeukenne:1976} for symmetric nuclear matter ).
Accordingly we obtain the {\it renormalized} $M_2$, which is approximately given by \cite{zuo:1999}:
\begin{equation}
\widetilde{M}_2^{\tau}(k,\omega)=
\sum_{\tau'} \left[ 1 - \kappa^{\tau'} \right] M_2^{\tau\tau'}(k,\omega) \ .
\label{e:Uren2}
\end{equation}
In terms of the off-energy-shell mass operator, one can readily calculate the neutron and proton
momentum distributions in asymmetric nuclear matter below and above the corresponding
Fermi momentum~\cite{jeukenne:1976,baldo:1991}:
\begin{equation}
n^{\tau}(k)=1+ \left[ \frac{\partial {\widetilde{U}^{\tau}_1(k,\omega)}} {\partial \omega}
\right]_{\omega=\epsilon^{\tau}(k)}, \ \ \ \ \ \  {\rm for} \ k < k^{\tau}_F
\end{equation}
\begin{equation}
n^{\tau}(k)= - \left[ \frac{\partial {\widetilde{U}^{\tau}_2(k,\omega)}} {\partial \omega}
\right]_{\omega=\epsilon^{\tau}(k)}, \ \ \ \ \ \  {\rm for} \ k> k^{\tau}_F
\end{equation}
where $\widetilde{U}^{\tau}_1$ and $\widetilde{U}^{\tau}_2$ denote the real parts of
$\widetilde{M}^{\tau}_1$ and $\widetilde{M}^{\tau}_2$, respectively.

\section{Results and Discussions}

\begin{figure}[tbh]
\begin{center}
\includegraphics[width=0.5\textwidth]{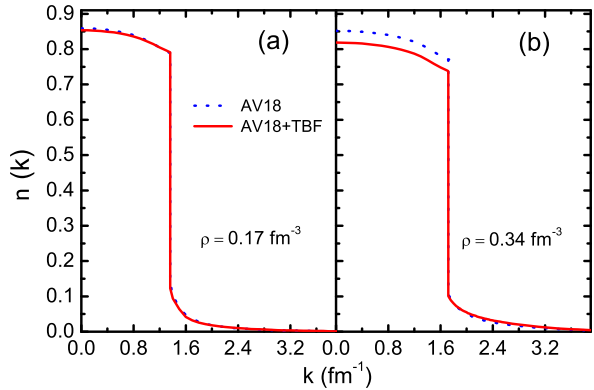}
\end{center}
\caption{(Color online) TBF Effect on the nucleon momentum distribution in symmetric nuclear matter ($\beta=0$)
for two densities 0.17 fm$^{-3}$ (left panel) and 0.34 fm$^{-3}$ (right panel).}\label{fig1}
\end{figure}

In Fig. \ref{fig1} we display the TBF effect on the predicted momentum distributions below and above the
corresponding Fermi momenta in symmetric nuclear matter ($\beta=0$) at two typical
densities $\rho=0.17$ and 0.34 fm$^{-3}$, respectively.
In the figure, the solid lines
correspond to the results obtained by including the TBF; the dashed ones are calculated by adopting
purely the $AV18$ two-body force alone.
It is clear from Fig.~\ref{fig1} that, due to the many-body correlations induced
by the $NN$ interaction, the s.p. states below the Fermi surface are partly empty, and those above the Fermi surface can be
partly occupied in the correlated ground state of nuclear matter. In the case of excluding the TBF, the
density dependence of the momentum distribution, as a function of the ratio $k/k_F$, is shown to be quite weak in the
density region considered here,
which is in good agreement with the previous EBHF calculation in Ref.~\cite{baldo:1991} by adopting the separable $AV14$
interaction and the prediction reported in Ref. \cite{rios:2009} by using the Green function method.
One may notice that
the TBF effect is negligibly small at low densities around and below the empirical saturation density $\rho_0=0.17$ fm$^{-3}$ of
nuclear matter. This is consistent with the conclusion in Ref. \cite{fantoni:1984} within
the correlated basis function approach by adopting the Urbana $v14$ interaction plus an effective three-body
interaction. At the high density $\rho=0.34$ fm$^{-3}$ which is well above the saturation density,
the TBF effect turns out to become noticeable.
By comparing the solid curves and the corresponding dashed curves in Fig. \ref{fig1},
it is seen that the TBF effect is to enhance the depletion of the momentum distribution
below the Fermi momentum at high densities, i.e., to reduce the occupation probability of the hole states.
This is readily understood since inclusion of the TBF is expected to induce
stronger short-range correlations in dense nuclear medium as compared with the case of excluding the TBF.
In both cases of including and excluding the TBF, the obtained
depletions of the hole states at zero momentum $(k=0)$ are roughly 15\% around the saturation density,
compatible with the previous predictions reported in Refs. \cite{grange:1987,baldo:1991,fantoni:1984,benhar:1990,rios:2009}.
Since the depletion of the s.p. hole states well below the Fermi momentum in nuclear matter
can be identified with the depletion of the occupation of the deeply bound s.p. levels in finite
nuclei \cite{ramos:1989,benhar:1990,muther:1995}, the present obtained
depletion at saturation density is also consistent with the experimental result in Ref. \cite{batenburg:2001}.

\begin{figure}[tbh]
\begin{center}
\includegraphics[width=0.5\textwidth]{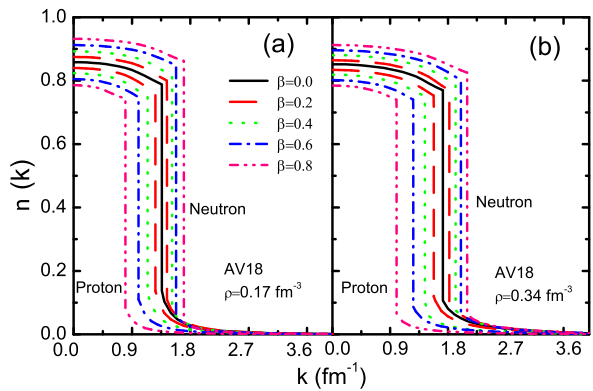}
\end{center}
\caption{(Color online) Neutron and proton momentum distributions in asymmetric nuclear matter at various asymmetries
for two densities 0.17 fm$^{-3}$ (left panel) and 0.34 fm$^{-3}$ (right panel).
The results are calculated without including the TBF.} \label{fig2}
\end{figure}

The isospin $T=0$ tensor component of the $NN$ interaction is expected to be crucial for
 determining the isospin dependence of the equation of state and s.p. properties of asymmetric
 nuclear matter \cite{bombaci:1991,Dieperink:2003,zuo:1999,zuo:2005}. The study of the neutron and proton momentum distributions as well as
 their isospin-asymmetry dependence may be helpful for understanding the properties of the short-range and tensor
 correlations in nuclear many-body systems \cite{subedi:2008,rios:2009}. In Fig. \ref{fig2} we report the neutron
 and proton momentum distributions below and above their respective Fermi momenta
in asymmetric nuclear matter at various asymmetries $\beta =0$, 0.2, 0.4, 0.6, and 0.8 for two typical densities
$\rho=0.17$ and 0.34 fm$^{-3}$, respectively. In the figure, the results are obtained
by adopting purely the $AV18$ two-body interaction and the TBF is not included.
It is clearly seen that the neutron and proton momentum distributions in asymmetric nuclear matter are different from their
common distribution in symmetric nuclear matter and the depletions of the neutron and proton hole states depend sensitively on the isospin-asymmetry.
As the isospin-asymmetry $\beta$ increases, the occupation probability of the neutron hole states below the neutron Fermi sea
becomes larger while the occupation of the proton hole states
gets smaller with respect to their common values in symmetric nuclear matter; that is, increasing the asymmetry leads
to a reduction of the depletion of the neutron hole states, while it enhances the depletion of the proton hole states.
The above result has also been found in Ref.~\cite{rios:2009} within the framework of the Green function approach.
Such an isospin-asymmetry dependence of the neutron and proton momentum distributions in asymmetric nuclear matter
implies that at a higher asymmetry, the effect of the short-range correlations, induced by the $NN$ interaction, becomes stronger (weaker)
on protons (neutrons), and can be understood according to the isospin-asymmetry dependence of the effect of the tensor
component of the $NN$ interaction in asymmetric nuclear matter.
As is well known, the isospin $T=0$ $SD$ tensor component may induce a strong
short-range correlation in nuclear medium and it plays a dominant role in determining the isospin vector parts of the properties
of asymmetric nuclear matter~\cite{zuo:1999,zuo:2005,baldo:1989,vidana:2011}. As the neutron excess increases,
the effect of the $T=0$ $SD$ tensor channel on protons (neutrons) from the surrounding neutrons (protons)
becomes stronger (weaker). Accordingly the $SD$ tensor component is expected to induce a larger (smaller) depletion of the
proton (neutron) Fermi sea at a higher asymmetry.
The definite evidence for the strong enhancement of the $np$ short-range correlations
over the $pp$ and $nn$ correlations observed at JLab~\cite{subedi:2008} has provided an experimental indication for the dominant role played
by the short-range tensor components of the $NN$ interactions in generating the $NN$ correlations~\cite{schiavilla:2007}.
The importance of the tensor force in determining the isospin-asymmetry dependence of the neutron and proton momentum distributions in asymmetric
nuclear matter has also been confirmed in Ref.~\cite{rios:2009} where the authors have calculated the depletion of the nuclear Fermi sea by adopting various
$NN$ interactions and they find that the iso-depletion [defined as the difference of the neutron and proton occupation
of the lowest momentum state in asymmetric nuclear matter, i.e., $n^n(k=0)-n^{p}(k=0)$] obtained by the $AV4'$ potential which has no tensor component is much lower than the predictions
by the other $NN$ interactions including the tensor components, especially at suprasaturation densities.

\begin{figure}[tbh]
\begin{center}
\includegraphics[width=0.5\textwidth]{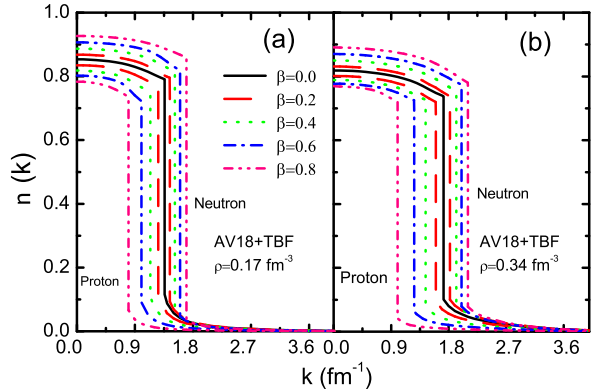}
\end{center}
\caption{ (Color online) The same as Fig.~\ref{fig2}, but the results are obtained by including the TBF. } \label{fig3}
\end{figure}

In Fig.~\ref{fig3}, we show the neutron and proton momentum distributions below and above their respective Fermi momenta,
predicted by adopting the $AV18$ interaction plus the TBF,
in asymmetric nuclear matter at various asymmetries $\beta =0$, 0.2, 0.4, 0.6, and 0.8 for two typical densities
$\rho=0.17$ and 0.34 fm$^{-3}$, respectively.
It is seen that, after including the TBF in the calculation, the predicted isospin-asymmetry dependence
of the neutron and proton momentum distributions remain the same as that obtained by adopting purely
the $AV18$ two-body force. By comparing the corresponding curves in Fig.~\ref{fig2} and Fig.~\ref{fig3},
one may notice that the TBF effect is negligibly small at sub-saturation densities.  At densities well above the
saturation density, the TBF leads
to an overall enhancement of the depletions of both the neutron and proton distributions below
their respective Fermi momenta since it may generate extra short-range $NN$ correlations which become strong
enough at high densities.

\begin{figure}[tbh]
\begin{center}
\includegraphics[width=0.5\textwidth]{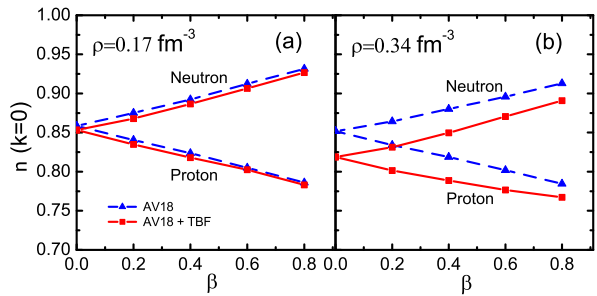}
\end{center}
\caption{(Color online) Neutron and proton momentum distributions at zero momentum in asymmetric nuclear matter
vs. isospin asymmetry $\beta$ for two densities 0.17 fm$^{-3}$ (left panel) and 0.34 fm$^{-3}$ (right panel).
The results are obtained for the two cases of including the TBF (solid curves) and excluding the TBF (dashed curves).} \label{fig4}
\end{figure}

In order to see more clearly the isospin dependence and the TBF effect,
in Fig.~\ref{fig4} we display the proton and neutron momentum distributions at zero momentum $k=0$ as functions of
the isospin-asymmetry $\beta$ in the two cases with (solid curves) and without (dashed curves) including the TBF.
It is clearly seen that the proton (neutron) occupation of the zero momentum state decreases (increases) almost
linearly as a function of asymmetry, indicating the short-range and tensor correlations become stronger (weaker) for
protons (neutrons) at a higher asymmetry in neutron-rich nuclear matter. A similar result has been reported
in Ref.~\cite{muther:1995} for the averaged neutron and proton occupation probabilities below their respective Fermi
seas at the empirical saturation density. The above predicted isospin splitting [i.e., $n^n(k=0)>n^p(k=0)$] of the
neutron and proton occupation of the lowest states in neutron-rich nuclear matter has also been found
in Ref.~\cite{rios:2009} by using the Green function approach. In Ref.~\cite{rios:2009}, it is shown that
at finite temperature the thermal effect may destroy the linear dependence of $n^n(k=0)$ and $n^p(k=0)$ on asymmetry $\beta$
at high enough asymmetries.
As expected, around and below the saturation densities, the TBF effect on the proton and neutron momentum distributions is
negligibly small in agreement with the result of Ref.~\cite{fantoni:1984}. However, at high densities ( for example $\rho=0.34$ fm$^{-3}$ )
well above the saturation density,
the TBF may affect sizably the neutron and proton occupations of their lowest states.
Inclusion of the TBF leads to an overall enhancement of the depletion of the neutron and proton hole states.
It is noticed from Fig.~\ref{fig4} that the TBF has almost no effect on the
iso-depletion $n^n(0)-n^{p}(0)$ up to $\rho=0.34$fm$^{-3}$,
which provides a support for the result of Ref.~\cite{rios:2009} where it has been shown that
once the tensor components fitting the experimental phase shifts are included,
various modern $NN$ interactions lead to almost the same iso-depletion although the momentum distributions
of neutrons and protons predicted by those different $NN$ interactions can be
quite different.

Before our summary, we shall give a brief discussion about the possible relevance of
the present results for the dense and highly asymmetric nuclear matter in the interior of neutron stars,
especially the
cooling of neutron stars via neutrino emission and the nucleon pairing in neutron stars.
Let us consider a neutron star consisting of neutrons, protons and electrons at $\beta$-equilibrium, i.e., a ($n,p,e$) neutron star model.
For a ($n,p,e$) neutron star, there are two different kinds of URCA processes for neutrino emission.
One is the direct URCA process, the other is the modified URCA process. The direct URCA process may lead to a much faster cooling of neutron stars than the
modified URCA process.
Under the assumption of a free Fermi momentum distribution of nucleon (i.e., the Fermi sea is fully occupied and the particle
states above the Fermi momentum are completely empty), the direct URCA process is only allowed if the
proton/neutron ratio $x\equiv\rho_p/\rho_n$ is greater than a threshold value of $x_c=1/8$ [which corresponds to a proton fraction of
$Y_p\equiv \rho_p / (\rho_n + \rho_p) =1/9$ and
an isospin-asymmetry of $\beta=7/9\simeq 0.8$]
in order to guarantee the momentum conservation~\cite{lattimer:1991}. The strong depletion of the proton Fermi sea and the partial occupation of the proton states
well above the Fermi momentum, induced by the short-range $np$ correlations in dense and highly asymmetric nuclear matter, may affect considerably the direct
URCA process as has been discussed in detail by Frankfurt {\it et al.} in Ref.~\cite{frankfurt:2008} where it is shown that the modification of the proton
momentum distribution in neutron star matter due to the short-range correlations leads to a significant enhancement of the neutrino luminosity of the
direct URCA process for temperatures much less than 1 MeV, and the direct URCA process may even have probability to occur for a small value of $x<0.1$.
Nucleon pairing in asymmetric nuclear
matter plays an important role in determining the cooling rate of neutron stars~\cite{gusakov:2005} and its strength has been shown to be rather sensitive to
the medium effect induced by nucleon-nucleon correlations~\cite{lombardo:2001}.
The large depletion of the proton Fermi sea in highly asymmetric nuclear matter due to the $T=0$ $np$ short-range correlations,
obtained in the present calculation, tends to reduce considerably the proton pairing in neutron stars~\cite{frankfurt:2008}. Another interesting topic
is the thermal transport parameters in dense asymmetric nuclear matter which have special
importance for the damping of nonradial modes of neutron stars and may depend sensitively on the depletion of nucleon
distribution~\cite{abrikov:1959,shternin:2008,zhang:2010,gusakov:2010}.
In general, the transport parameters,
including the shear viscosity and thermal conductivity, may
be calculated within the framework of the Landau Fermi liquid approach in which a free Fermi momentum distribution is assumed
as the equilibrium distribution~\cite{abrikov:1959}. Therefore, a strong depleted proton Fermi sea in dense and supra-dense asymmetric nuclear matter
is expected to affect substantially the transport properties in the interior of neutron stars.

\section{Summary}

In summary, we have investigated the TBF effect on the proton and neutron momentum distributions in asymmetric nuclear
matter within the framework of the EBHF approach by adopting the $AV18$ two-body interaction supplemented with a microscopic TBF.
In symmetric nuclear matter ($\beta=0$), the obtained depletion of the hole states deep inside the Fermi sea is roughly $15\%$ at the empirical
saturation density, in agreement with the previous predictions~\cite{grange:1987,baldo:1991,fantoni:1984,benhar:1990,rios:2009}
and the experimental value in Ref.~\cite{batenburg:2001}.
In asymmetric nuclear matter ($\beta>0$), the neutron and proton momentum distributions turn out to become
different and may split with respect to their common distribution in symmetric nuclear matter.
It is shown that increasing the isospin-asymmetry $\beta$ tends to enhance the depletion of the proton Fermi sea while it
reduces the depletion of the neutron Fermi sea, which implies that at a higher asymmetry, the effect of the tensor correlations
 induced by the $NN$ interaction may become stronger on protons while it gets weaker on neutrons.
At zero momentum, the neutron occupation probability increases while the proton occupation
  decreases almost linearly as a function of asymmetry.
The present obtained isospin dependence of the neutron and proton momentum distributions in asymmetric nuclear matter
is in good agreement with the recent prediction in Ref.~\cite{rios:2009} within the framework of the Green function method.
At low densities around and below the nuclear saturation density, the TBF effect on the predicted momentum distributions
is found to be negligibly weak in agreement with the conclusion of Ref.~\cite{fantoni:1984}. At high densities well
above the saturation density, the TBF is expected to induce strong enough extra short-range correlations and its effect turns out
to become noticeable. In dense asymmetric nuclear matter, inclusion of the TBF effect may lead to
an overall enhancement of both the depletion of the neutron and proton Fermi seas for all the asymmetries considered.
Although the TBF affects sizably the neutron and proton momentum distributions at high densities well above the saturation density,
its effect on
the iso-depletion of the nuclear Fermi sea (i.e., the difference of the neutron and proton occupation probabilities)
in asymmetric nuclear matter is shown to be quite small in the density region up to two times saturation density.
The present results are expect to have significant implication for the cooling, nucleon pairing and transport properties of neutron stars.

\section*{Acknowledgments}

{The work is supported by the National Natural Science
Foundation of China (11175219,10875151,10740420550), the Major State Basic
Research Developing Program of China under Grant No. 2013CB834405, the
Knowledge Innovation Project (KJCX2-EW-N01) of Chinese Academy of
Sciences, the Chinese Academy of Sciences visiting professorship for senior international
scientists (Grant No. 2009J2-26), the Project of Knowledge Innovation Program
(PKIP) of Chinese Academy of Sciences, Grant No. KJCX2.YW.W10, and the CAS/SAFEA International Partnership Program for Creative
Research Teams (CXTD-J2005-1).}

\end{CJK*}
\end{document}